# Phase-Modulated Interferometry, Spectroscopy, and Refractometry using Entangled Photon Pairs

*Jonathan Lavoie,*[1,3] *Tiemo Landes,*[1,3] *Amr Tamimi,*[2,3] *Brian J. Smith,*[1,3] *Andrew H. Marcus,*[2,3,*] *and Michael G. Raymer*[1,3,*]

[1.] Department of Physics, University of Oregon, Eugene, OR 97403, USA

[2.] Dept of Chemistry and Biochemistry, University of Oregon, Eugene, OR 97403, USA

[3.] Oregon Center for Optical, Molecular & Quantum Science, University of Oregon, Eugene, OR 97403, USA

Corresponding authors: ahmarcus@uoregon.edu, raymer@uoregon.edu

**Abstract**

The authors demonstrate a form of two-photon-counting interferometry by measuring the coincidence counts between single-photon-counting detectors at an output port of a Mach-Zehnder Interferometer (MZI) following injection of broad-band time-frequency-entangled photon pairs (EPP) generated from collinear spontaneous parametric down conversion into a single input port. Spectroscopy and refractometry are performed on a sample inserted in one internal path of the MZI by scanning the other path in length, which acquires phase and amplitude information about the sample's linear response. Phase modulation and lock-in detection are introduced to increase detection signal-to-noise ratio and implement a 'down-sampling' technique for scanning the interferometer delay, which reduces the sampling requirements needed to reproduce fully the temporal interference pattern. The phase-modulation technique also allows the contributions of various quantum-state pathways leading to the final detection outcomes to be extracted individually. Feynman diagrams frequently used in the context of molecular spectroscopy are used to describe the interferences resulting from the coherence properties of time-frequency EPPs passing through the MZI. These results are an important step toward implementation of a proposed method for molecular spectroscopy—quantum-light-enhanced two-dimensional spectroscopy.

# 1. Introduction

Two-photon interference is a core phenomenon in quantum optics and a key resource in quantum information science.[1] Measurements that determine the correlations between properties of pairs of photons play a central role in fundamental tests of physics,[2-4] and in practical applications.[5-9] Time-frequency entangled photon pairs (EPP) generated by spontaneous parametric down-conversion can be tightly correlated in time while also anti-correlated in frequency such that the sum of the photon energies is sharply defined, analogous to the Einstein-Podolsky-Rosen (EPR) state, yielding non-classical correlations in quantum optical measurements.[10, 11]

It has been recognized previously that the quantum correlations inherent in time-frequency EPP might be exploited to enhance the measurement performance of nonlinear spectroscopy and metrology. For example, time-frequency EPP were used for quantum optical coherence tomography (QOCT)[12, 13] and for enhanced two-photon absorption (TPA).[14, 15] The established benefits of using EPP in these experiments were: 1) increased TPA cross sections, 2) linear scaling of TPA rates with photon fluence,[16-18] 3) increased signal-to-noise ratio (SNR) in optical metrology, and 4) elimination of thermal background noise.[19-21] In imaging applications, broad-band time-frequency EPP, combined with Hong-Ou-Mandel (HOM) fourth-order interference,[1] was used to achieve approximately 20 micrometers axial resolution in strongly scattering media.[13] Moreover, it was demonstrated that EPP could be used to achieve simultaneously narrow frequency- and narrow time-response in the TPA of an atomic vapor.[22]

While the present paper focuses on linear-response spectroscopy and refractometry using EPP in coincidence detection, we also look ahead toward implementing nonlinear spectroscopy using EPP. Recent proposals have suggested new ways to exploit time-frequency correlations in EPP to enhance nonlinear spectroscopic signals measured in multi-dimensional optical spectroscopy,[23-26] which normally utilizes sequences of ultrafast laser pulses with inter-pulse delays controlled by adjustable path interferometers.[27] Our group proposed entangled-photon-pair two-dimensional fluorescence spectroscopy (EPP-2DFS), in which the coherent laser pulses are replaced by time-frequency EPP.[26]

Here we characterize the properties of EPP transmitted through a single Mach-Zehnder interferometer (MZI) containing an internal linear-response medium in one MZI arm, using phase-sensitive lock-in detection, analogous to the methods used in the "classical" 2DFS scheme developed previously by Marcus and co-workers.[28, 29] In contrast to standard Fourier-transform infrared spectroscopy (FTIR), in which the sample is placed after the MZI exit port,[30] the placement of the sample internal to the interferometer – as in interferometric refractometry – allows both absorptive and dispersive responses of the sample to be measured.[31, 32] This enables characterizing the dispersion of a transparent medium, such as quartz, for which there is no absorptive component in the scanned spectral region.

An important aspect of our approach to EPP coincidence detection is the use of a 'down-sampling' technique that permits scanning the interferometer delay with larger step sizes than required by the standard methods that are typically much smaller than an optical wavelength, thereby increasing the efficiency of data collection. The resulting speedup in data collection efficiency is important, for example, when contemplating a full two- (or three-) dimensional spectrum, as is done in multi-dimensional spectroscopy.[35] The present results are thus a significant step toward the full implementation of nonlinear EPP-2DFS.

A key aspect of the phase-modulation method is that it enables separation of several known quantum optical effects that occur simultaneously—one-photon interference, Hong-Ou-Mandel (HOM) two-photon interference,[1] and bi-photon "N00N-state" behavior.[4,33,34] The HOM-like process occurs when the photon pair is split into separate paths at the first beam splitter of the MZI, and the N00N-like process occurs when the photon pair travels together in one arm or the other. The separation of these MZI processes is analogous to the isolation of distinct material state quantum Liouville pathways in classical multidimensional laser spectroscopy. Such experiments are described using double-sided Feynman diagrams, which are a powerful technique for understanding molecular structure and dynamics. Furthermore, our analysis shows that interference between HOM-like processes and N00N-like processes leads to an effective 'one-photon' interaction with the internal sample medium, giving an absorptive signal similar to FTIR.

## 2. Experimental Scheme

**Figure 1** shows a schematic of the experimental apparatus. The optical source is a beam of broad-band, time-frequency entangled photon pairs, with entanglement coherence time in the femtosecond regime. The short coherence time is comparable to the relaxation time of a molecular absorber (or ring-down time of a spectral notch filter) with line width on the order of tens-of-nanometers in wavelength.

The EPP are generated by pumping a collinear spontaneous parametric down conversion (SPDC) process in a 1 mm-thick beta barium borate (BBO) crystal (labeled $\chi^{(2)}$) using a narrow-band continuous-wave (CW) laser. The pump laser is a frequency-doubled solid-state laser, with single-frequency output at 266 nm. The SPDC phase-matching condition is Type-1, so that the signal and idler polarizations are parallel to each other and orthogonal to that of the pump. The phase-matching condition is adjusted by varying the crystal angle such that the EPP occupy a common spatial and polarization mode, and the EPP spectrum is degenerate with signal and idler both centered at 532 nm. The EPP bandwidth is approximately 35 nm. The average power of the 266 nm pump beam is maintained at approximately 75 mW, and the rate of EPP generation is approximately 150 kHz per mW of pump power.

A low-pass optical filter (LP) blocks the pump beam, and the SPDC light is coupled into a single-mode optical fiber for spatial-mode filtering before it is directed to the entrance 50/50 beam-splitter (BS) of a Mach-Zehnder interferometer (MZI). In the upper arm of the MZI is an adjustable delay ($\tau$) and in the lower arm is a linear-response medium ($\eta(\omega)$), which may be alternatively removed.

Each arm of the MZI contains an acousto-optic modulator (AOM), which applies a constant-rate linear phase sweep (constant frequency shift) at frequency $\nu_1$ = 110.000 MHz or $\nu_1$ = 110.020 MHz, creating a relative time-varying phase $\Delta\phi(t) = 2\pi\nu_{21}t$, where $v_{21} = v_2 - v_1$ = 20 kHz. The beams within the two arms of the MZI are recombined at the exit 50/50 BS. One of the exiting beams is divided using a second 50/50 BS, and photon pairs are monitored using Geiger-mode avalanche photo-diodes (APDs) as single-photon detectors ($D_I$ and $D_{II}$) and a coincidence detection circuit (coinc).

To implement the down-sampling technique used in "classical" multidimensional spectroscopy,[28, 29, 36] a coherent reference laser is used to probe the instantaneous relative phase of the MZI. This is accomplished by back-propagating a helium-neon (HeNe) laser with frequency $\omega_R$ (wavelength 633 nm) through the MZI, which is monitored by a photodiode (PD) at the unused port of the entrance BS. The sinusoidal voltage from the PD provides an analog reference for phase-sensitive detection of the coincidence signal using a standard lock-in amplifier.[28, 29]

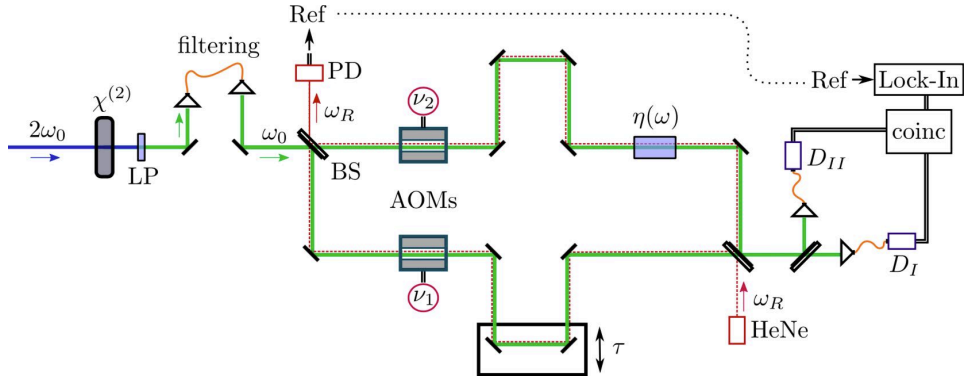

**Figure 1.** Two-photon phase-modulated interferometer. $\chi^{(2)}$: BBO crystal, LP: low-pass spectral filter, filtering: single-mode fiber, BS: beam splitter, PD: photo-diode, AOM: acousto-optic modulator, $\eta(\omega)$: linear response of sample, $D_I$ and $D_{II}$: single mode-fiber-coupled avalanche photodetectors.

The setup can be used to perform photon coincidence measurements in two different data acquisition modes. First, with the AOMs removed and using "small" (~ 15 nm) steps to scan the relative delay of the MZI, we can obtain a "fully-sampled" interferometric scan. Alternatively, it is possible to implement a phase-sensitive detection scheme, which uses the lock-in amplifier to measure the component of the coincidence signal that oscillates in time ($t$) at particular harmonics of the reference frequency. By using lock-in detection, the signal oscillates as a function of delay $\tau$ at a significantly lower frequency than that of the optical carrier of the EPP beam.

As mentioned previously, the relative phase between the MZI arms is continuously swept at the AOM modulation frequency $v_{21} = \Delta\phi(t)/2\pi t$. For a fixed delay $\tau$, the rate of photon detection events (coincidence or singles) is the sum of three contributions, which we call "0f," "1f" and "2f", denoting their dependence on the AOM phase sweep $\Delta\phi(t)$:

$$A(t,\tau) = A_{0f} + 2\,\text{Re}\left\{ A_{1f} \exp\left[i\left(\Delta\phi(t) - \omega_0\tau - \theta_{1f}\right)\right] + A_{2f} \exp\left[i\left(2\Delta\phi(t) - 2\omega_0\tau - \theta_{2f}\right)\right]\right\} \quad (1)$$

In this equation, $\omega_0$ is the center frequency of the EPP, the amplitudes $A_{0f}, A_{1f}$ and $A_{2f}$ are real, corresponding to complex amplitudes of the 1f and 2f components $A_{1f}\exp(-i\theta_{1f})$ and $A_{2f}\exp(-i\theta_{2f})$, respectively. We discuss the theoretical origins of the 0f, 1f and 2f terms below. We note here that the presence of the 2f signal is due to the fact that a bi-photon state has twice the energy of a one-photon state, such that the quantum phase of the bi-photon state accumulates at twice the frequency of the one-photon state.[4,33,34]

The voltage from the photodiode that monitors the HeNe reference, with carrier frequency $\omega_R$, is proportional to $\cos[\Delta\phi(t)-\omega_R\tau]$. Operationally, the lock-in amplifier multiplies the signal by a waveform constructed from the reference voltage: $\cos[m(\Delta\phi(t)-\omega_R\tau)]$, or $\sin[m(\Delta\phi(t)-\omega_R\tau)]$, where $m$ is an integer. Then it applies a low-pass filter, which is equivalent to averaging uniformly over the phase from 0 to $2\pi$. For example, the 1f in-phase and in-quadrature components are extracted as:

$$X_{1f}(\tau) \equiv \left\langle A(t,\tau)\cos\left[\Delta\phi(t)-\omega_R\tau\right]\right\rangle_{\text{low-pass}} = A_{1f}\cos\left[(\omega_0-\omega_R)\tau+\theta_{1f}\right]$$
$$Y_{1f}(\tau) \equiv -\left\langle A(t,\tau)\sin\left[\Delta\phi(t)-\omega_R\tau\right]\right\rangle_{\text{low-pass}} = -A_{1f}\sin\left[(\omega_0-\omega_R)\tau+\theta_{1f}\right]$$
(2)

We note that the low-pass-filtered signal varies with the MZI delay $\tau$ at the down-sampled frequency $\omega_0-\omega_R$, which is a much lower frequency than that of either the optical carrier or the reference laser, provided that $\omega_0$ and $\omega_R$ are not too dissimilar. The method therefore requires fewer steps in $\tau$ to obtain the information contained by the interferogram. Moreover, the down-sampled signal is more stable, in comparison to the fully-sampled signal, with respect to uncontrolled fluctuations of the relative MZI phase $\Delta\phi$ or delay $\tau$.[28, 29]

By combining the in-phase and in-quadrature components of the 1f signal, we may construct the complex-valued signal $Z_{1f}(\tau)=X_{1f}(\tau)+iY_{1f}(\tau)$, or determine the corresponding amplitude and phase:

$$A_{1f} = \left[X_{1f}^2(\tau)+Y_{1f}^2(\tau)\right]^{1/2}, \quad \theta_{1f} = \tan^{-1}\left[Y_{1f}(\tau)/X_{1f}(\tau)\right]$$
(3)

We apply a similar procedure to extract the 2f component of the signal. In this case, the lock-in amplifier multiplies the signal by $\cos[2\Delta\phi(t)-2\omega_R\tau]$ or $-\sin[2\Delta\phi(t)-2\omega_R\tau]$, after which a low-pass filter is applied. We thus obtain the in-phase and in-quadrature components of the 2f signal, $X_{2f}(\tau)$ and $Y_{2f}(\tau)$, which we combine to determine the amplitude and phase, $A_{2f}$ and $\theta_{2f}$, respectively. Finally, the 0f component $A_{0f}$ is obtained by simply recording the time-averaged (i.e., phase-averaged) signal at a given MZI step delay τ, effectively washing out the fast oscillations leaving the DC component.

Among the advantages of this method are an increased SNR resulting from the use of phase-sensitive (lock-in) detection, and a reduction of the number of MZI steps needed to obtain the spectroscopic information contained by the interferogram. These benefits carry over from the phase-modulation techniques developed for "classical" 2DFS.[28, 29]

Several related experiments have been carried out previously. Boitier et al used a direct two-photon detector to time resolve counts from a two-photon Michelson interferometer.[37] Kalashnikov et al used a Type-2 SPDC interferometer to reconstruct the absorption spectrum of a partially transmitting sample.[38]

### 3. Conceptual Model

In **Figure 2(a)**, we show a simple depiction of a pair of indistinguishable photons entering the MZI in a common input port (labeled A) and exiting in a common output port (labeled D). The upper and lower paths of the MZI are labeled "1" and "2", respectively, and a sample (colored blue) may be optionally inserted into path 2. Figures 2(b) and 2(c) show both photons traveling in a common path of the MZI, and Figure 2(d) shows one photon in each path. The quantum amplitudes for all four of the two-photon interference pathways add coherently to yield the final amplitude for two photons to emerge at the output port.

We consider first the case without a sample. In general, varying the delay while measuring photon coincidences provides a novel way to reveal information about two-photon interference. Specifically, Figure 2(b) and 2(c) depict the so-called N00N state $|2,0\rangle + |0,2\rangle$ inside the interferometer, where $|n,m\rangle$ corresponds to $n$ ($m$) photons in the upper (lower) path, and for simplicity we have ignored the time-frequency entanglement in the notation. It is well established that such a state when emerging from the output port would display interference "fringes" oscillating at twice the frequency (half the wavelength) expected in a classical or one-photon interference experiment.[4,33,34,39] In contrast, the state depicted in Figure 2(d) is $|1,1\rangle$. This $|1,1\rangle$ component of the state leads to Hong-Ou-Mandel (HOM) interference—that is, two photons in one or the other output port.[1] As we discuss below, both N00N and HOM types of two-photon interferences contribute coherently to the final output amplitude of the MZI.

When a sample is present in path 2, measurements of two-photon coincidences as a function of the MZI delay τ provide spectroscopic information about the sample, in a fashion similar to Fourier-transform refractometry.[30, 31, 36] This information is provided by the interactions between the sample and the four different two-photon interference pathways. For example, Figure 2(b) shows both photons interacting with the sample, while Figure 2(c) shows neither photon interacting with the sample, and Figure 2(d) shows one photon interacting with the sample and the other not. All of these possibilities combine coherently to yield the final amplitudes and probabilities for coincidences.

In addition to describing the phase-modulated EPP-MZI scheme for observing quantum optical effects, a major goal of this study is to define fundamentally the information obtainable from a material system using this scheme. Closely related studies include those by Kalashnikov, which used non-collinear EPP, separated by propagation direction from the source crystal, to probe interferometrically the absorption of a sample placed within one path of the MZI.[37]

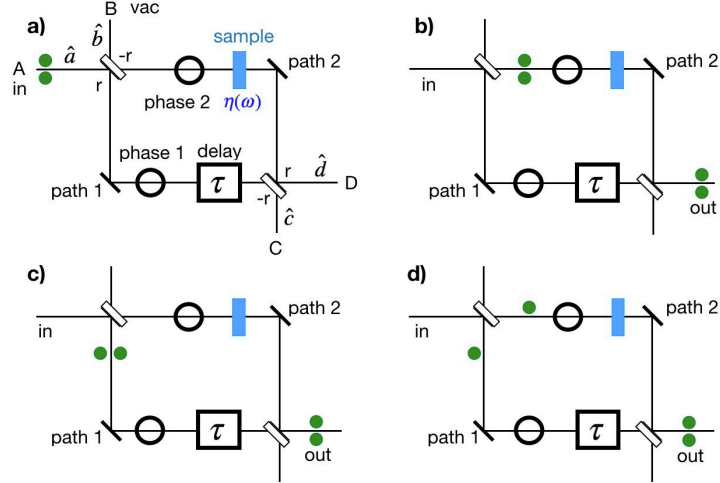

**Figure 2.** Possible processes in a two-photon interferometer with: a) two indistinguishable photons incident in the "*A*" port, leading to the output state with two photons exiting the "*D*" port. A superposition of b) and c) corresponds to the N00N state. d) corresponds to the HOM state.

## 4. Theoretical Model

Here we detail the derivation of the coincidence rates, both with and without a linear-response medium placed in one path of the MZI. Referring to Figure 2(a), the MZI is comprised of two beam splitters (BS), each with real amplitude transmissivity, *t,* and reflectivity, ±r, with phase conventions for reflectivity as shown in the figure. The input signal port is denoted *A*, with annihilation operator $\hat{a}$. The other input port (*B*) is vacuum. Output ports are denoted *C* or *D*, with annihilation operators $\hat{c}$ or $\hat{d}$, respectively.

The two-photon component of the input state arising from collinear type-I down conversion is [40-43]

$$|\psi\rangle = \int d\omega \int d\tilde{\omega}\, \psi(\omega, \tilde{\omega})\, \hat{a}^\dagger(\omega)\, \hat{a}^\dagger(\tilde{\omega}) |vac\rangle \tag{4}$$

where the form of the joint-spectral amplitude (JSA) $\psi(\omega, \tilde{\omega})$ depends on the pump laser spectrum and the phase-matching conditions. Because the type-I source does not distinguish the two photons, the JSA is necessarily symmetric under label exchange, $\psi(\tilde{\omega}, \omega) = \psi(\omega, \tilde{\omega})$. The normalized time-dependent electric-field operators that are detected in coincidence are denoted:

$$\hat{E}_j^{(+)}(t) = \int d\omega \, e^{-i\omega t} \hat{e}_j(\omega) \tag{5}$$

Where $\hat{e}_1 = \hat{c}$, $\hat{e}_2 = \hat{d}$, etc., such that $\left[\hat{e}_i(\tilde{\omega}), \hat{e}_j^\dagger(\omega)\right] = \delta(\tilde{\omega} - \omega)\delta_{ij}$. The instantaneous coincidence detection probability for detectors at output ports *j* and *k* is proportional to:

$$C_{jk}(t_j, t_k) = \left\langle \hat{E}_j^{(-)}(t_j)\hat{E}_k^{(-)}(t_k)\hat{E}_j^{(+)}(t_j)\hat{E}_k^{(+)}(t_k) \right\rangle \tag{6}$$

A coincidence of two photons in a single output port is modeled by taking *j* = *k*. The detectors integrate over a time window $2T_W$ that is long compared to the sub-picosecond temporal characteristics of the field, so the observed coincidence rate is proportional to

$$\begin{aligned} R_{jk} &= \int_{-T_W}^{T_W} dt_j \int_{-T_W}^{T_W} dt_k \, C_{jk}(t_j, t_k) \\ &\simeq N \int d\omega \int d\tilde{\omega} \left\langle \hat{e}_j^\dagger(\omega)\hat{e}_k^\dagger(\tilde{\omega})\hat{e}_j(\omega)\hat{e}_k(\tilde{\omega}) \right\rangle \end{aligned} \tag{7}$$

In Equation (7), the frequency integrals extend over the entire range of the field spectrum, and *N* is a normalization constant. Because we consider fields with at most two photons in a detection interval, the rate can be written (by inserting the identity operator, in which only the vacuum state contributes) as

$$R_{jk} = N \int d\omega \int d\tilde{\omega} \left| f_{jk}(\omega, \tilde{\omega}) \right|^2 \tag{8}$$

where

$$f_{jk}(\omega, \tilde{\omega}) = \left\langle vac \left| \hat{e}_j(\omega)\hat{e}_k(\tilde{\omega}) \right| \psi \right\rangle \tag{9}$$

is the corresponding two-photon detection amplitude, whose square modulus is the joint detection probability.

Forward propagation through the MZI is represented by the transformation

$$\begin{aligned} \hat{c}(\omega) &= g_{ca}(\omega)\hat{a}(\omega) + g_{cb}(\omega)\hat{b}(\omega) \\ \hat{d}(\omega) &= g_{da}(\omega)\hat{a}(\omega) + g_{db}(\omega)\hat{b}(\omega) \end{aligned} \tag{10}$$

where

$$\begin{aligned} g_{da}(\omega) &= rt\,\eta(\omega)e^{i\phi_2} + rte^{i\omega\tau}e^{i\phi_1} \quad,\quad g_{db}(\omega) = -r^2\eta(\omega)e^{i\phi_2} + t^2 e^{i\omega\tau}e^{i\phi_1} \\ g_{ca}(\omega) &= t^2\eta(\omega)e^{i\phi_2} - r^2 e^{i\omega\tau}e^{i\phi_1} \quad,\quad g_{cb}(\omega) = -rt\,\eta(\omega)e^{i\phi_2} - rte^{i\omega\tau}e^{i\phi_1} \end{aligned} \tag{11}$$

If the linear response function of the sample η(ω) has an imaginary part, the transformation described by Equation (10) and (11) is not unitary, and a correct treatment would require additional operators to represent loss channels. Here we are concerned only with the two-photon coincidence signal, which is unaffected by absorption processes of the sample. We can therefore ignore the effects of loss channels in our current treatment. With this observation, we write the two-photon detection amplitude

$$f_{jk}(\omega,\tilde{\omega}) = g_{ja}(\omega)g_{ka}(\tilde{\omega})\langle vac|\hat{a}(\omega)\hat{a}(\tilde{\omega})|\psi\rangle \tag{12}$$

The JSA is found, using Equation(4) and Equation(12), to be automatically symmetrized via commutation relations:

$$\langle vac|\hat{a}(\omega)\hat{a}(\tilde{\omega})|\psi\rangle = \psi(\omega,\tilde{\omega}) + \psi(\tilde{\omega},\omega) \equiv \psi_{SYM}(\omega,\tilde{\omega}) \tag{13}$$

such that $\psi_{SYM}(\tilde{\omega},\omega) = \psi_{SYM}(\omega,\tilde{\omega})$, which is consistent with the fact that photons are bosons.

In our case, the laser that pumps the down conversion crystal is nearly monochromatic with frequency $\omega_p$, such that the generated photon pairs are strongly anti-correlated in frequency, with a spectral distribution $S(\omega)$ obeying the condition

$$|\psi_{SYM}(\omega,\tilde{\omega})|^2 = S(\omega)\delta(\omega_p - \omega - \tilde{\omega}) \tag{14}$$

Thus, the coincidence rate, from Equation(8), is

$$R_{jk} = N\int d\omega |g_{ja}(\omega)g_{ka}(\tilde{\omega} = \omega_p - \omega)|^2 S(\omega) \tag{15}$$

Note that the coincidence rate described by Equation(15) depends on the spectrum of the EPP, but not on its spectral phase. Therefore, the spectral phase of the EPP cannot be inferred from MZI-based coincidence measurements. Nevertheless, the measurements do reveal the spectral phase of the medium placed in one arm of the interferometer, as we demonstrate below.

We derive the rate of detecting two photons in the output beam *D* of the MZI [see Figure 2(a)], which is proportional to the coincidence rate between detectors $D_I$ and $D_{II}$:

$$R_{dd} = N\int d\omega |g_{da}(\omega)g_{da}(\tilde{\omega} = \omega_p - \omega)|^2 S(\omega) \tag{16}$$

The MZI transfer function $|g_{da}(\omega)g_{da}(\tilde{\omega})|^2$ appearing in Equation(16) can be expanded to identify the specific time-evolved quantum-state amplitude, and the associated phase signature, of each contributing term to the coincidence rate:

$$\left| g_{da}(\omega) g_{da}(\tilde{\omega}) \right|^2 = \left[ rt\eta(\omega)e^{i\phi_2} + rte^{i\omega\tau}e^{i\phi_1} \right]\left[ rt\eta(\tilde{\omega})e^{i\phi_2} + rte^{i\tilde{\omega}\tau}e^{i\phi_1} \right]^2 \qquad (17)$$

$$= \left| \; i_H \; + \; ii_H \; + \; iii_N \; + \; iv_N \; \right|^2$$

The four terms in this equation are denoted

$$i_H = r^2 t^2 \eta(\omega) e^{i\tilde{\omega}\tau} e^{i(\phi_1+\phi_2)} \;,\; ii_H = r^2 t^2 \eta(\tilde{\omega}) e^{i\omega\tau} e^{i(\phi_1+\phi_2)}$$
$$iii_N = r^2 t^2 \eta(\omega)\eta(\tilde{\omega}) e^{i2\phi_2} \;,\; iv_N = r^2 t^2 e^{i(\omega+\tilde{\omega})\tau} e^{i2\phi_1} \qquad (18)$$

For the case with no sample inside the MZI, the four terms of Equation(18) correspond to the Feynman diagrams shown in Figure 3, which describe the different quantum pathways that the photon pairs in beam *A* are transmitted by the MZI to the output beam *D*. The subscripts *H* and *N* label each diagram as related to the evolution of a Hong-Ou-Mandel (HOM) pathway or a N00N pathway. For example, diagram $i_H$ and $ii_H$ describe the evolution of HOM pathways, in which the photon pair is initially in the tensor product state $|AA\rangle$ before it is split by the entrance beam splitter (BS1) into separate paths (*rt*). Immediately after the entrance beam splitter, the photons are in the state $|p1p2\rangle$ or $|p2p1\rangle$, such that one photon evolves in path 1 and the other in path 2. Each photon passes through a phase shifter, thus the state acquires the phase $(\phi_1 + \phi_2)$. At the exit beam splitter (BS2), one photon reflects (*r*) while the other undergoes a time delay (denoted $\omega\tau$ or $\tilde{\omega}\tau$), leading to the intermediate state $|p1D\rangle$ or $|Dp1\rangle$. After the delay, the remaining photon transmits (*t*), leading to the final state $|DD\rangle$ (both photons in path *D*). For the cases described by diagrams $iii_N$ and $iv_N$, both photons go into a common path (*tt or rr*) at the entrance beam splitter, thus giving the N00N pathways $|p1p1\rangle$ and $|p2p2\rangle$. Both photons pass through a common phase shifter, thus the state acquires the phase $2\phi_1$ or $2\phi_2$. Finally, either both of the photons undergo a time delay, denoted $(\omega+\tilde{\omega})\tau$, or neither does, such that the N00N pathways have no intermediate state with one of the two photons delayed inside the MZI while the other is outside. At the exit beam splitter, both photons reflect (*rr*) or both transmit (*tt*), again leading to the final state $|DD\rangle$.

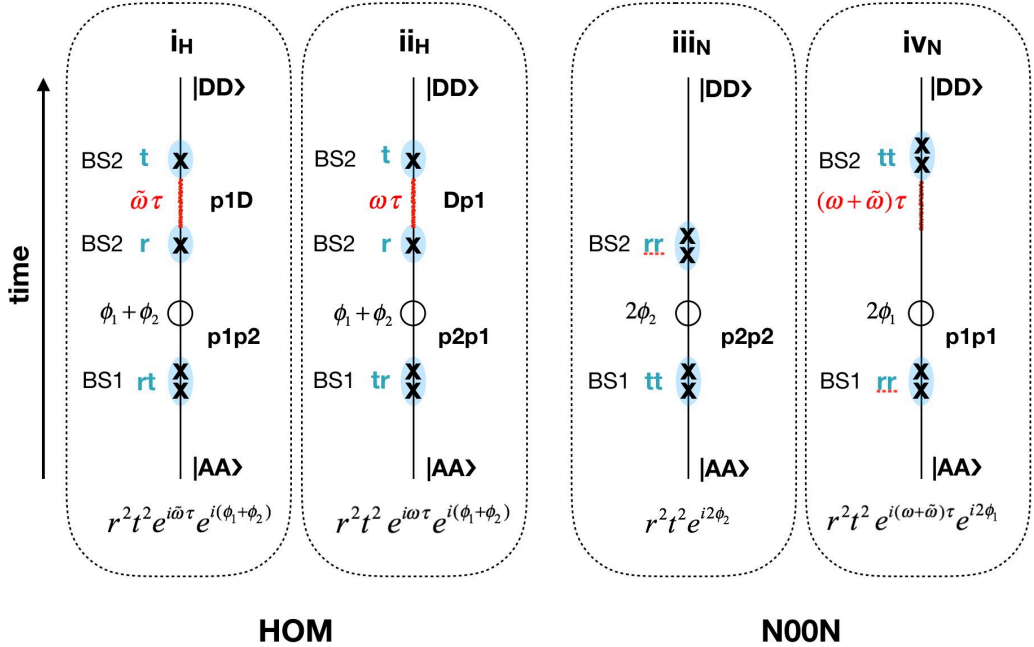

**Figure 3.** Feynman diagrams in the case that there is *no sample* in the MZI. Time runs from bottom to top, with the states labeled on the right and processes labeled on the left. State labels: AA = two photons in input path A, p1=path 1, p2=path 2, DD = two photons in output path D.

As indicated by Equation (16) and (17), the two-photon coincidence rate is the spectrally integrated square-modulus of the sum of diagrams $i_H$ through $iv_N$. By slowly sweeping the MZI phases and demodulating the coincidence rate (as described in the Experimental section above), distinct quantum Liouville pathway contributions to the final outcome can be isolated according to their characteristic phase signatures.

These pathways can be visualized with double-sided Feynman diagrams, as is often done in nonlinear optics and multidimensional spectroscopy.[44, 45] In **Figure 4**, each diagram represents an outer product between one of the terms in Equation(18) and the Hermitian conjugate of another. In this representation, vertical paths on the left side of the diagram are kets, while those on the right side are bras.

We first consider the double-sided diagrams with the "1f" phase signature $\pm(\phi_2 - \phi_1)$. Recall that the relative phase of the MZI is swept: $\Delta\phi(t) = \phi_2 - \phi_1 = 2\pi\nu_{21}t$, where $\nu_{21}$=20 kHz, which is slow in comparison to relevant molecular time scales (1 fs – 1 ns), but fast in comparison to environmental noise (∼ 1 ms). From Equation(17), the "1f" terms are:

$$i_H \cdot iv_N^* + ii_H \cdot iv_N^* + iii_N \cdot i_H^* + iii_N \cdot ii_H^* + cc \qquad (19)$$

The coherent sum of terms given by this equation determines the intensity of the 1f component of the two-photon coincidence rate, after demodulation by the lock-in amplifier. Each of these diagrams describes the time evolution of an element of the density operator as the photon pair undergoes successive transitions from the initial population $|AA\rangle\langle AA|$ to the final population $|DD\rangle\langle DD|$. The intervening transitions involve coherences between HOM and N00N pathways. For example, the double-sided diagram "$i_H$-$iv_N$" describes a pathway in which the initial population is first converted to the coherence $|p1p2\rangle\langle p1p1|$, and next to the coherence $|p1D\rangle\langle p1p1|$ before undergoing a final transition to create population in the $D$ beam.

The double-sided diagrams offer an intuitive description of the quantum interferences in the MZI. For example, diagram "$i_H$-$iv_N$" shows that the initial population in AA is transformed to a coherence between a N00N state with both photons in the long path (path 1) and a HOM state with one of the two photons in the long path (path1) and the other HOM photon having already left the MZI. Thus, the difference of phase accumulation between these two state components varies as $\omega\tau$, so it contributes to the 1f signal. Similar descriptions apply to the other three diagrams in Figure 4.

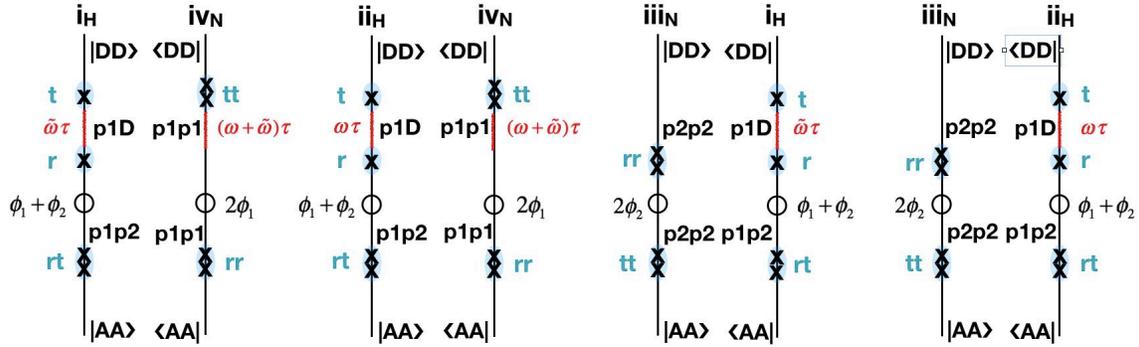

**Figure 4.** The four double-sided diagrams that contribute to the 1f phase signature. Time runs from bottom to top. State labels are indicated between the vertical time lines, and processes affecting kets (on the left line) and bras (on the right line) are indicated outside the lines.

For convenience of notation, we use the normalized rates, $\tilde{R}_{dd} = R_{dd}/2r^4t^4N$ in the following calculations, with $R_{dd}$ defined in Equation (16). The result of integrating the sum of the four 1f diagrams (using the state symmetry and normalization) is:

$$\tilde{R}_{dd(1f)}(t,\tau) = 2\int d\omega \, \text{Re}\left[\eta(\omega)\left\{\left|\eta(\omega_p - \omega)\right|^2 + 1\right\}e^{-i(\omega\tau - \Delta\phi(t))}\right]S(\omega) \qquad (20)$$

where the phase difference was defined earlier as $\Delta\phi(t) = 2\pi v_{21}t$. Equation (20) relates the 1f coincidence rate to a convolution of the sample susceptibility η(ω) with the EPP spectrum $S(\omega)$, which is a function of the MZI delay τ. For a fixed delay, the effect of the linear phase sweep is to modulate the coincidence rate at the 1f frequency ($v_{21}$ = 20 kHz).

We next consider the double-sided diagrams that have the 2f phase signature $\pm 2(\phi_2 - \phi_1)$. From Equation(18), there are two such terms:

$$iii_N \cdot iv_N^* + cc \tag{21}$$

which are illustrated in **Figure 5**. As we discussed above, in N00N processes both photons pass together through path 1 or path 2, thus acquiring twice the phase that would be otherwise picked up by a single photon passing through that path of the MZI.

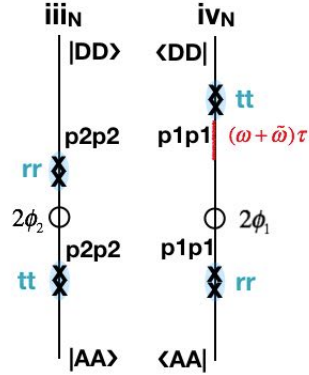

**Figure 5.** The double-sided diagram that leads to the "2f" phase signature.

Corresponding to the diagrams of Figure 5, we calculate the normalized coincidence rate with 2f phase signature

$$\tilde{R}_{dd(2f)}(t,\tau) = \text{Re}\left[ e^{-i(2\omega_0 \tau - 2\Delta\phi)} \int d\omega\, \eta(\omega) \eta(2\omega_0 - \omega) S(\omega) \right] \tag{22}$$

where we have used the fact that the pump frequency is equal to twice the EEP carrier frequency: $\omega_p = 2\omega_0$. We note that for fixed delay τ, the effect of the linear phase sweep, $\Delta\phi = 2\pi v_{21} t$, is to modulate the coincidence rate at the 2f frequency (2 $v_{21}$ = 40 kHz).

Calculations and diagrams analogous to those shown above for 1f and 2f lead to the phase-independent coincidence rate, designated as 0f. There are six different products of terms with the 0f phase signature

$$i_H \cdot i_H^* + ii_H \cdot ii_H^* + iii_N \cdot iii_N^* + iv_N \cdot iv_N^* + \left( i_H \cdot ii_H^* + cc \right) \tag{23}$$

The resulting 0f coincidence detection rate is:

$$\tilde{R}_{dd(0f)}(\tau) = \frac{1}{2}\int d\omega \left\{ |\eta(\omega)|^2 + |\eta(2\omega_0 - \omega)|^2 + |\eta(\omega)\eta(2\omega_0 - \omega)|^2 + 1 \right\} S(\omega) \quad (24)$$
$$+ \int d\omega \, \text{Re}\left[ \eta(\omega)\eta^*(2\omega_0 - \omega) e^{i(2\omega_0 - \omega)\tau} \right] S(\omega)$$

The 0f component of the signal described by Equation(24) contains partial information about material dispersion, consistent with well-known dispersion-cancellation effects occurring in HOM interference experiments. That is, a Taylor expansion of the propagation constant around the central frequency $\omega_0$ results in cancellation of all even-order dispersive terms, while retaining the odd-order terms.[46, 47]

There are two experimental approaches to measure the two photon coincidence signals described by Equation (20), (22) and (24). The first is to run the experiment without the AOMs, this requires fine step sizes (e.g. 15 nm) and results in a 'fully-sampled' scan, which is the sum of all three terms.

The second measures the three phase signatures 0f, 1f, and 2f separately using the lock-in phase-sensitive detection scheme, as described earlier. The 0f component of the signal $A_{0f}$ [see Equation(1)] is extracted from the two-photon coincidence rate by simply recording the time-averaged signal given by Equation(24) at a given step delay, without the use of the lock-in amplifier:

$$A_{0f} = \tilde{R}_{dd(0f)} \quad (25)$$

We obtain the complex-valued down-sampled 1f coincidence rate, $Z_{dd(1f)}(\tau) = X_{dd(1f)}(\tau) + iY_{dd(1f)}(\tau)$ by multiplying $\tilde{R}_{dd(1f)}$ given by Equation(20) by the phase factor $\exp[-i(\Delta\phi - \omega_R\tau)]$, where $\omega_R$ is the frequency of the reference laser, and applying the low-pass filter operation described by Equation(2):

$$Z_{dd(1f)}(\tau) = \int d\omega \, \eta(\omega)\left\{ 1 + |\eta(2\omega_0 - \omega)|^2 \right\} e^{-i(\omega - \omega_R)\tau} S(\omega) \quad (26)$$

where we have again used the fact that $\omega_p = 2\omega_0$. We see from this equation that the demodulation method simply replaces the term $2\text{Re}[\eta(\omega)e^{-i(\omega\tau - \Delta\phi)}]$ in Equation(20) by the term $\eta(\omega)e^{-i(\omega - \omega_R)\tau}$. Because $S(\omega)$ is peaked at the signal central frequency $\omega_0$, which is not very different from $\omega_R$, the 1f demodulated signal varies slowly as $\tau$ is scanned, roughly as $\exp[-i(\omega_0 - \omega_R)\tau]$. This variation is much slower than that of the fully-sampled scan, illustrating that down-sampling allows for the collection of fewer data points to acquire the same information present in the fully-sampled scan. We note that the in-phase and in-quadrature signals, $X_{dd(1f)}(\tau)$ and $Y_{dd(1f)}(\tau)$, contain full information about the spectral phase $\theta_{dd(1f)}$ and amplitude $A_{dd(1f)}$ of the material susceptibility $\eta(\omega)$, which are recovered by the experiment according to the relations given by Equation(3).

We obtain the complex-valued down-sampled 2f coincidence rate by multiplying $\tilde{R}_{dd(2f)}$ [Equation (22)] by the phase factor $\exp[-i(2\Delta\phi - 2\omega_R \tau)]$ and applying the low-pass filter operation

$$Z_{dd(2f)}(\tau) = \frac{1}{2} e^{-i2(\omega_0 - \omega_R)\tau} \int d\omega\, \eta(\omega)\eta(2\omega_0 - \omega) S(\omega) \tag{27}$$

This equation shows that the down-sampled 2f signal oscillates for all values of the MZI delay at the frequency $2(\omega_0 - \omega_R)$. Since the spectral phase of $\eta(\omega)$ is integrated in Equation(27), its effect on the 2f signal is simply an overall phase-shift of the oscillations in $Z_{dd(2f)}(\tau)$. We further note that the spectral phase of the incident EPP light does not enter into any of the coincidence rates.

For completeness, we include here the two photon coincidence rates for the case in which there is no sample within the MZI. In this case, the modulated coincidence rates are:

$$\begin{aligned}
\tilde{R}_{dd(0f)}(\tau) &= 2\int d\omega S(\omega) + \text{Re}\left[\int d\omega\, e^{i(2\omega_0 - 2\omega)\tau} S(\omega)\right] \\
\tilde{R}_{dd(1f)}(t,\tau) &= 4\,\text{Re}\left[\int d\omega\, e^{-i(\omega\tau - \Delta\phi(t))} S(\omega)\right] \\
\tilde{R}_{dd(2f)}(t,\tau) &= \cos(2\omega_0 \tau - 2\Delta\phi(t)) \int d\omega S(\omega)
\end{aligned} \tag{28}$$

The corresponding down-sampled coincidence rates are:

$$\begin{aligned}
A_{dd(0f)}(\tau) &= 2\int d\omega S(\omega) + \text{Re}\left[\int d\omega\, e^{i(2\omega_0 - 2\omega)\tau} S(\omega)\right] \\
Z_{dd(1f)}(\tau) &= 2\int d\omega\, e^{-i(\omega - \omega_R)\tau} S(\omega) \\
Z_{dd(2f)}(\tau) &= \frac{1}{2} e^{-i2(\omega_0 - \omega_R)\tau} \int d\omega\, S(\omega)
\end{aligned} \tag{29}$$

We note that another consequence of the down-sampling method is to reduce the amplitude of the 1f and 2f components by a factor (1/2), while that of the 0f component is unaffected.

## 4.1 Model Calculations with a Gaussian EPP Spectrum

To test our treatment of EPP phase-modulated interferometry, we initially performed studies using broad-band EPP light without a sample in the MZI. We first perform model calculations in which the EPP source spectrum is approximated as a normalized Gaussian function,

$$S(\omega) = (\pi\alpha^2)^{-1/2} \exp[-(\omega-\omega_0)^2/\alpha^2] \qquad (30)$$

where $\omega_0 = \omega_p/2$ and α determines the full-width-at-half-maximum (FWHM) of the EPP spectrum (FWHM = $2\alpha\sqrt{\ln(2)}$). Assuming no loss or dispersion in the MZI, the integrals given by Equation(28) can be evaluated explicitly to yield the fully-sampled coincidence rates.

$$\begin{aligned}
\tilde{R}_{dd(0f)}(\tau) &= 2 + e^{-\alpha^2\tau^2} \\
\tilde{R}_{dd(1f)}(t,\tau) &= 4e^{-\frac{1}{4}\alpha^2\tau^2} \cos(\Delta\phi(t) - \omega_0\tau) \\
\tilde{R}_{dd(2f)}(t,\tau) &= \cos[2\Delta\phi(t) - 2\omega_0\tau]
\end{aligned} \qquad (31)$$

The corresponding down-sampled coincidence rates are:

$$\begin{aligned}
A_{dd(0f)}(\tau) &= 2 + e^{-\alpha^2\tau^2} \\
Z_{dd(1f)}(\tau) &= 2e^{-\frac{1}{4}\alpha^2\tau^2} e^{-i(\omega_0-\omega_R)\tau} \\
Z_{dd(2f)}(\tau) &= \frac{1}{2} e^{-i2(\omega_0-\omega_R)\tau}
\end{aligned} \qquad (32)$$

In **Figure 6**, we present calculations of the various coincidence rates using a Gaussian-modeled EPP spectrum with FWHM = 35 nm [Figure 6(a)], and the 0f, 1f and 2f down-sampled components [Equation(32)] using a Gaussian FWHM = 15 nm [Figures. 6(b) – 6(c) respectively]. These values for the FWHM were chosen to match the parameters used in the experiments discussed below.

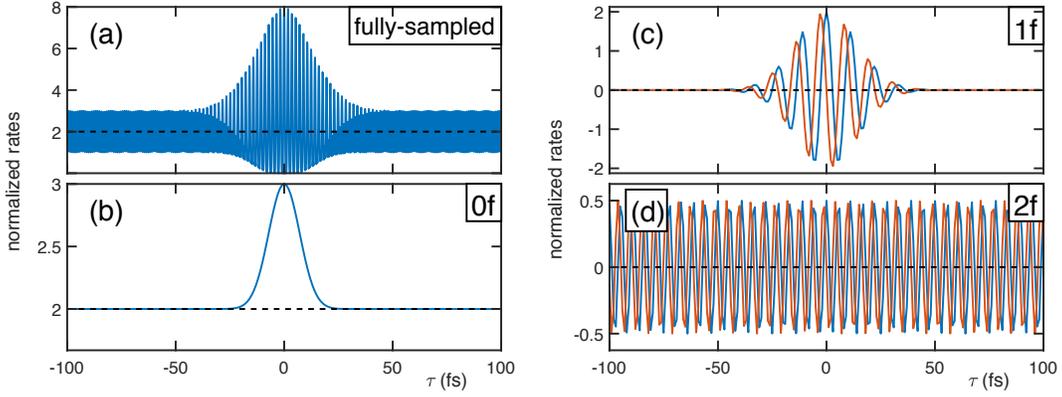

**Figure 6.** Simulated two-photon coincidence rate interference in the time domain. (a) Fully-sampled interferogram using a 35 nm FWHM EPP spectrum; (b) 0f down-sampled component using FWHM = 15 nm; (c) 1f down-sampled component using FWHM = 15 nm; (d) 2f down-sampled component using FWHM = 15 nm. These FWHM values were chosen to match the values in the experimental measurements.

The fully-sampled interferogram shown in Figure 6(a) exhibits a combination of behaviors, each of which may be separately determined using the down-sampling method. A key point is that the down-sampled signals vary slowly with delay, reflecting the difference between the optical carrier frequency and that of the reference HeNe, whereas the fully-sampled rate reflects the optical carrier frequency alone. While it is possible to extract each of the down-sampled signals by Fourier transforming the fully-sampled total scan, we show below that the quality of the experimental signals is greatly enhanced using the down-sampling method.

Referring back to the conceptual model presented above, in Figure 6(b) we see evidence for a Hong-Ou-Mandel (HOM) peak in the rate $A_{dd(0f)}$. The HOM peak occurs, rather than a dip, because in our configuration we detect the probability for two photons to emerge from the same exit port, which is enhanced by the bosonic nature of photons.[1] In Figure 6(d) we see evidence for bi-photon (N00N-state) 2f oscillation, which is indicative of the energy of a bi-photon being twice that of a single photon. The N00N-state interference pattern persists for delays much longer than the coherence time of the light because it arises from cases where both photons take the upper path or both photons take the lower path, so that temporal overlap is not needed.[4, 33, 34, 39] In Figure 6(c) we see that the 1f component is the Fourier transform of the power spectrum of the incident light, as indicated by Equation (31) and (32). According to the standard Wiener-Khinchin theorem, this equals the two-time autocorrelation function of the field. All of these behaviors are combined in the fully-sampled scan in Figure 6(a). The phase-modulation method proposed here allows the three behaviors to be separated cleanly and detected efficiently, with high SNR.

## 5. Experimental Results

In **Figure 7**, we show the measured spectrum of the EPP with the BBO crystal rotated to the type-I collinear phase-matching angle. The small peaks on either side result from the typical phase-matching behavior as a result of material dispersion in the BBO. The FWHM of the Gaussian fit is 36.5 nm, or in angular frequency $\Delta\omega = 2\pi \times 3.87 \times 10^{13} \mathrm{s}^{-1}$.

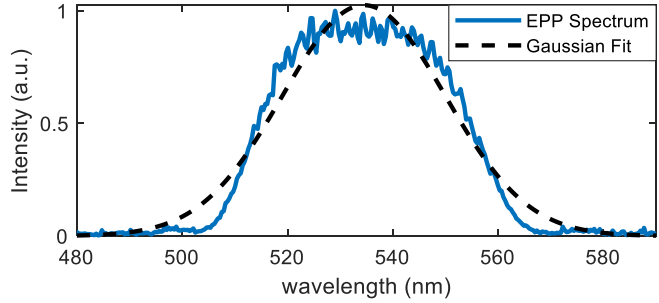

**Figure 7**. EPP spectrum coupled into a single-mode optical fiber before the Mach-Zehnder Interferometer. The dashed line shows a Gaussian model fit (Equation (30) with FWHM = 36.5 nm).

### 5.1. Experimental Results With No Sample

In **Figure 8**, we show the measured coincidence signals obtained using the experimental setup shown in Figure 1. For the case of the fully-sampled scan [Figure 8(a)], the AOMs were removed from the MZI paths. Under these conditions, the full EPP spectrum with FWHM = 35 nm could be coupled into the single-mode-fiber-coupled APDs. However, for the case of the down-sampled scans, the AOMs introduced angular dispersion of the beams, which were spectrally narrowed while coupling into the single-mode-fibers. We modeled the effect of this narrowing for the down-sampled scans by using an EPP spectrum with FWHM = 15 nm.

The results of our measurements are in excellent agreement with simulated interferograms (such as those shown in Figure 6) when allowing for the spectral narrowing and the presence of the interferometer noise and sampling statistics, which are not modeled in the theoretical predictions. These signals—HOM (0f), one-photon (1f) and N00N (2f) —were also reported in using a fully-sampled scan similar to that shown in Figure 7(a).[39] However, in that study the three types of signals were not separately measured or analyzed, as we have done using our new Fourier transform and down-sampling method.

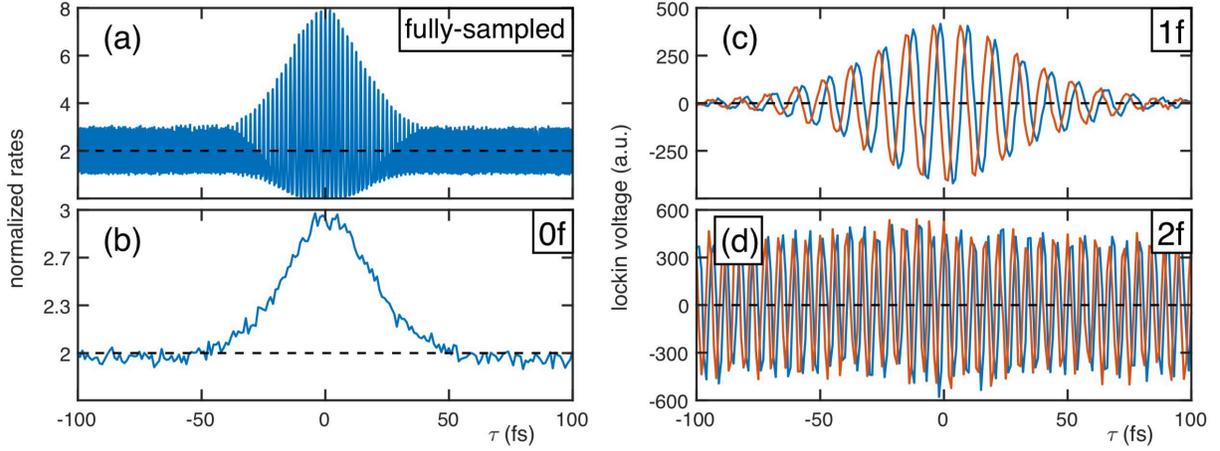

**Figure 8.** Experimental two-photon coincidence-rate interference in the time domain using the EPP spectrum shown in Figure 7. (a) Fully-sampled scan using 15 nm steps. The dashed boxes indicate regions of distinct behaviors, which are expanded in Figure 9; (b) phase-modulation is applied, and the (time-averaged) 0f signal is scanned using 150 nm steps; (c) the phase-modulated signal is demodulated by the lock-in amplifier at the 1f frequency (20 kHz), which is scanned using 150 nm steps; (d) the phase-modulated signal is demodulated at the 2f frequency (40 kHz), which is scanned using 150 nm steps.

In **Figure 9**, we show expanded views of the regions enclosed within the dashed boxes shown in Figure 8(a). In Figure 9(a), we show a section of the "N00N region" of the fully-sampled scan. The frequency of oscillations in this region is twice that of the EPP center frequency (i.e., $\omega_p = 2\omega_0$), where $\omega_p$ is the frequency of the narrow-band pump laser (with wavelength 266 nm) used to create the EPP spectrum with center wavelength 532 nm. In Figure 9(b), we show a section of the "transition region," where oscillations at the frequency $\omega_0$ grow in amplitude while oscillations at the frequency $2\omega_0$ diminish as the path difference of the MZI delay $\tau$ is adjusted to zero. In Figure 9(c), we show a section of the "balanced-MZI region" near $\tau = 0$, where the sum of all three terms in Equation 31 approaches $8\cos^4(\omega_0 \tau / 2)$ for constant $\Delta\phi(t)$. The five central fringes can be fit to this formula with high precision, indicating a very high visibility.

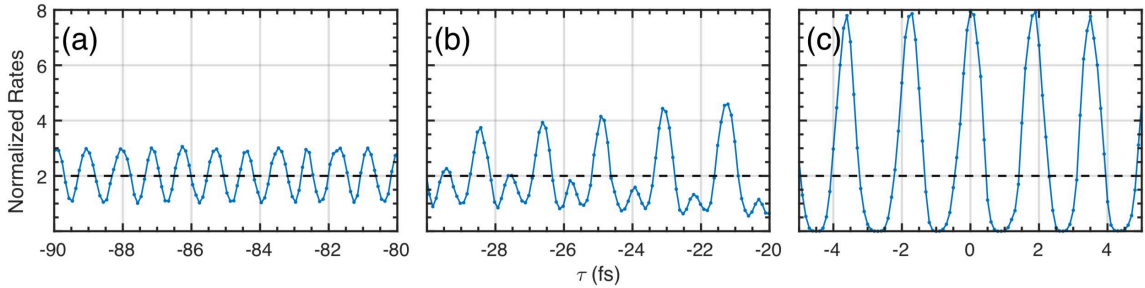

**Figure 9.** Expanded views of the fully-sampled experimental scan shown in Figure

8(a). (a) The "N00N region" exhibits oscillations at the pump frequency $\omega_p = 2\omega_0$ with wavelength 266 nm; (b) The "transition region" exhibits interference between oscillations at $\omega_0$ and $2\omega_0$; (c) The "balanced-MZI region" exhibits interference between oscillations at $\omega_0$ and $2\omega_0$, which approaches the form $8\cos^4(\omega_0\tau/2)$ near $\tau = 0$.

In **Figure 10**, we plot the Fourier transforms of the fully-scanned coincidence signal [Figure 10(a)] and of the 0f, 1f and 2f components of the coincidence signal [Figure 10(b)], which are shown in Figure 8. These data clearly show the well-separated components of the signal. The three peaks in Figure 10(b) have been positioned horizontally to correspond to the proper optical frequencies in the fully sampled plot. And they have each been normalized to have maximum value of 1. The phase-modulation method allows us to isolate these distinct features before analyzing the data.

A striking feature of the data in Figure 10 is the noise reduction, particularly in the 2f Fourier-transform peak of the down-sampled scan in comparison to that of the fully-sampled scan. The fully-sampled scan is conducted with a non-stabilized interferometer which therefore exhibits phase jitter due to optical path fluctuations (generally below a kilohertz), as well as slow phase drift over the course of the scan. This noise is particularly impactful in the 2f Fourier component due to the short wavelength (equivalent to that of the pump laser, 266 nm). In addition to the strongly reduced dependence of the phase on the MZI's path length due to the down-sampling in the demodulated experiments, the demodulation with a real-time optical reference largely neutralizes the effect of phase-jitter on the signal-to-noise ratio in the down-sampled case.

The noise reduction is clearly visible in comparing the noise baselines of Figure 10 (a) and (b), and the width of the 2f peak, which ideally reflects the very narrow linewidth of the CW pump laser. This also means that the line shapes in the fully-sampled scans are substantially convolved with the broadening due to the phase jitter and drift. The insensitivity to these effects in the down-sampled case means that meaningful line-shape analysis can be performed without needing to actively or passively stabilize the interferometer.

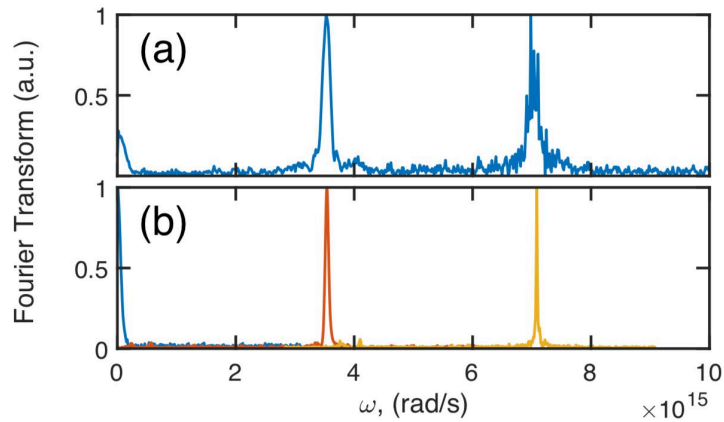

**Figure 10**. Fourier transforms of two-photon coincidence interference for (a) fully-sampled case and down-sampled data: (b) 0f (blue), 1f (red), and 2f

(yellow) data from Figure 8.

## 5.2. Experimental Results with Dispersive Sample

As an initial test of the response of the EPP-MZI apparatus to the presence of a material sample, we placed a 30.8 mm crystalline quartz slab – a transparent, dispersive material – into one arm of the MZI, as shown in Figure 1. For such a material, the linear transfer (transmission) function has the form:

$$\eta(\omega) = e^{ik(\omega)L} \tag{33}$$

where $L$ is the thickness of the material. We can expand the propagation constant around the central frequency and write $k(\omega) \approx \alpha(\omega - \omega_0) + \beta(\omega - \omega_0)^2$, where we have ignored higher order terms, and $\alpha = k'(\omega)\big|_{\omega_0}$ is the inverse group velocity and $\beta = (1/2)k''(\omega)\big|_{\omega_0}$ is one half the group-velocity dispersion (GVD). The group-delay dispersion (GDD) is $\text{GDD} = \beta L$. The 3.0-cm quartz slab was oriented so that the polarization of the EPP light was parallel to the ordinary crystal axis. The value of the GVD for quartz is well known to be 75.970 fs² mm⁻¹.[48]

In **Figure 11**, we show the results of our two-photon coincidence measurements in the presence of the quartz sample. For the case of the fully-sampled interferogram [Figure 11(a)], we observe interference (beating) between the 1f and 2f components of the signal that is qualitatively different from the pattern we observed in the absence of the sample. The phase-modulation method allows us to measure separately the 0f, 1f and 2f components. For example, we see that the effect of the sample is to introduce quadratic- and higher-order dispersion in the 1f component of the signal, which is shown in Figure 11(c). The 2f component [Figure 11(d)] has constant amplitude, and a carrier frequency equal to twice the difference between the reference and signal frequencies, as expected according to Equation(28). We note that the narrowness of the 0f 'HOM peak' in Figure 11(b) is due to the fact that the quadratic phase is canceled in Equation(24). It can be easily shown that all terms of the dispersive response that are symmetric around the center frequency (i.e., even powers in the Taylor expansion around $\omega_0$) cancel, while all odd powers of the dispersive response remain.[49]

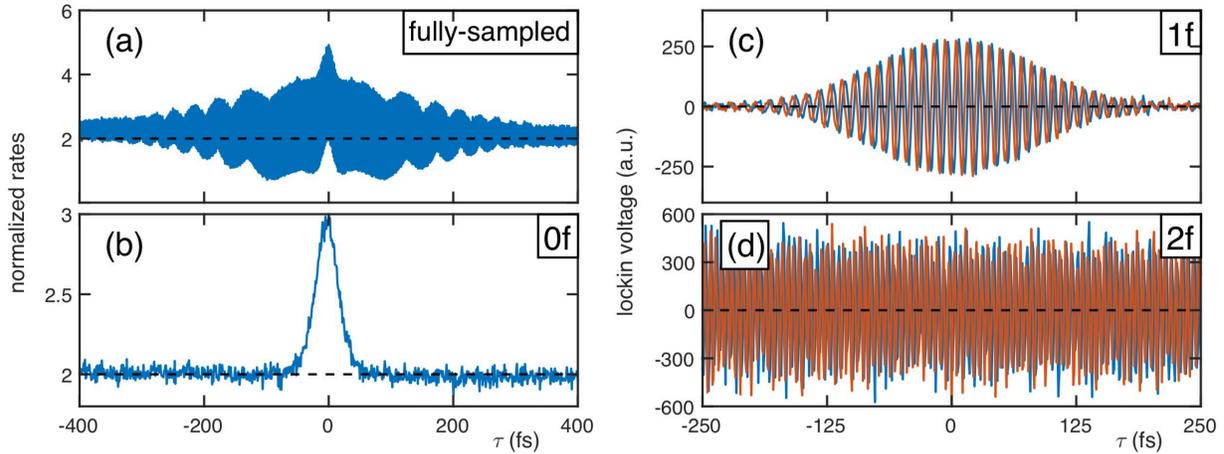

**Figure 11.** Measured two-photon coincidence rate interference with a 3-cm crystalline quartz slab inserted into one arm of the MZI (as shown in Figure 1). (a) Fully-sampled scan with 15 nm steps. Individual steps are not resolved in the plot; (b) 0f signal component with 150 nm steps; (c) 1f signal component with 150 nm steps; (d) 2f signal component with 150 nm steps.

As for the case with no sample, the signals measured with phase modulation were inadvertently spectrally narrowed at the single-mode fiber after the interferometer, which resulted in broadening of the MZI delay scans of the 0f interference. Perhaps counter-intuitively, the time-domain 1f envelope is narrower (roughly a factor of two) than in the fully-sampled case. This is due to the quadratic dispersion in the quartz sample. While a Fourier-limited pulse would be narrower in time, the quadratic phase broadens this pulse substantially given the larger bandwidth of the fully-sampled scan. The interplay of these two effects results in the observed envelope. This result is in good agreement with simulations (not shown).

In **Figure 12**, we show the Fourier transform of the 1f component, with the linear phase removed in the time domain.

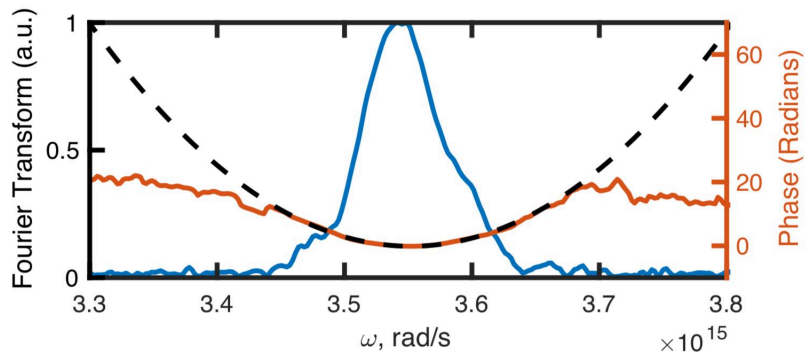

**Figure 12.** Quartz interference scans in frequency domain. (Linear phase removed in time-domain data.)

From the data shown in Figure 12, we determined the GVD using a quadratic fit weighted by the intensity of the magnitude spectrum. From this procedure, we obtained the value GVD = 71.7 ± 0.3 fs$^2$mm$^{-1}$, which is in good agreement with reported values (75.970 fs$^2$mm$^{-1}$) to within unknown systematic errors in the apparatus. To highlight the quadratic phase in this analysis, we removed the group delay from the time-domain data by centering the interferogram around zero delay. The resulting value of the GDD that we measured is in good agreement with reported values.

A Gaussian fit to the measured spectrum yields a FWHM of ~15 nm, which agrees well with the spectrum of the EPP at the output of the MZI, as measured by a conventional spectrometer.

### 5.3. Experimental Results with a Notch Filter inside the MZI

As a second test of the EPP-MZI method, we inserted into the sample arm of the MZI a notch filter, which strongly blocks transmission in a fixed spectral band and transmits light outside of this band. This represents a test of the method's ability to measure both absorptive and dispersive responses. In **Figure 13**, we show an overlay of the measured spectrum of the EPP source with and without the notch filter inserted, and our simulations of these spectra.

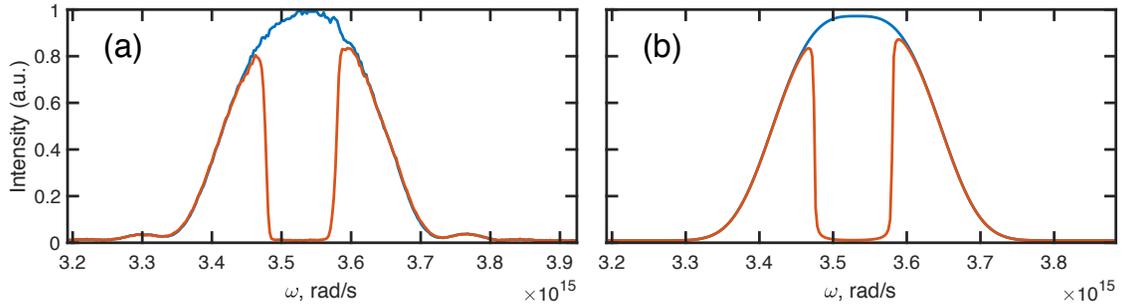

**Figure 13.** (a) Measured EPP Spectrum with (in red) and without (blue) notch filter inserted. (b) Modeled (super-Gaussian) EPP spectrum with and without notch filter.

We model the magnitude of the transmission function of the filter, $|\eta(\omega)|$, by the following:

$$|\eta(\omega)| = \frac{1}{2} + \frac{1}{\pi}\arctan\left[s\{(\omega - \omega_n)^2 - w^2\}\right] \quad (34)$$

where the full transfer function, $\eta(\omega) = |\eta(\omega)|\exp[i\phi(\omega)]$, includes a spectral phase, $\phi(\omega)$, determined by numerically computing the Kramers-Kronig relation for transmission and phase of a linear system, as given in.[50] Here $\omega_n$ is the center frequency, $w$ sets the width, and *s* sets the steepness of the notch. In **Figure 14**, we plot this function (in blue) using the parameters: $\omega_n$ = 3.527 ×10$^{15}$ rad s$^{-1}$, $w$ = 5.285 ×10$^{13}$ rad s$^{-1}$, and $s$ = 10$^{-16}$ s$^2$rad$^{-2}$.

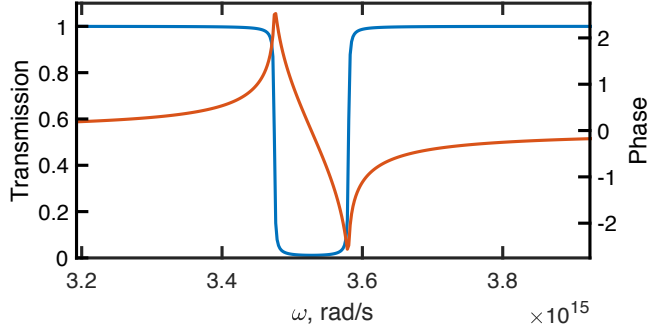

**Figure 14.** Modeled transmission $|\eta(\omega)|$, blue curve: and (dispersive part) phase, red curve, from Equation(34) and its Kramers-Kronig counterpart.

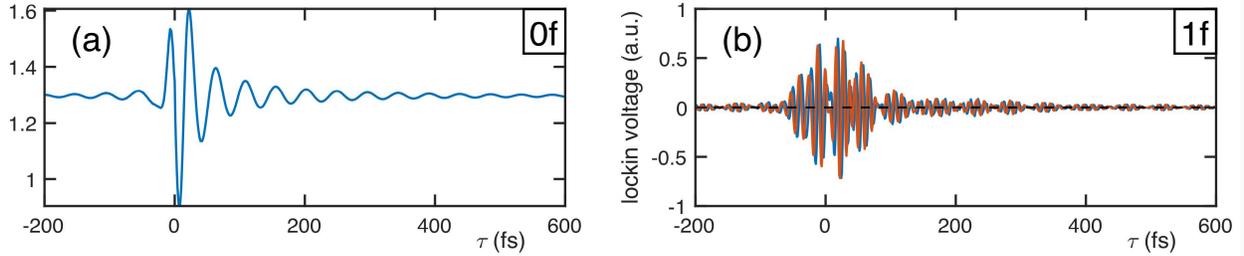

**Figure 15.** Simulated two-photon coincidence interference. (a) 0f HOM (b) 1f.

The simulations of the observed interference shown in **Figure 15** capture two important effects: ringing due to the absorption of the notch, and asymmetry in interferometer delay due to the causal nature of the filter's response. This information is contained in the spectral phase contributed by the notch filter.

For the time domain of the 0f component, the even-order dispersion cancellation effects we observed for the quartz sample do not hold for the notch filter. This is due to the fact that the HOM interference isolated in the 0f signal depends on both the absorptive response and on odd-orders of the dispersive response, which are not cancelled in HOM interference.[49] This effect can be easily derived from the theoretical prediction for the 0f component given by Equation(24), through the term $\eta(\omega)\eta^*(2\omega_0 - \omega)$. This in effect maintains information about the odd orders of dispersion, without the contributions of the even orders.

Another effect due to the absorptive part of the sample is seen in the term $\eta(\omega)\left\{1 + |\eta(2\omega_0 - \omega)|^2\right\}$ in Equation(20). While the phase information of the transfer function is maintained, unmodified, in the interference pattern the absorptive response is

modified by the square-magnitude of the transfer function as observed by the spectrally anti-correlated photon. The 0f absorptive response is similarly modified by the term: $\eta(\omega)\eta^*(2\omega_0 - \omega)$. These effects are reflected in the experimental results shown in **Figure 16**, which are in general agreement with the simulations in Figure 15.

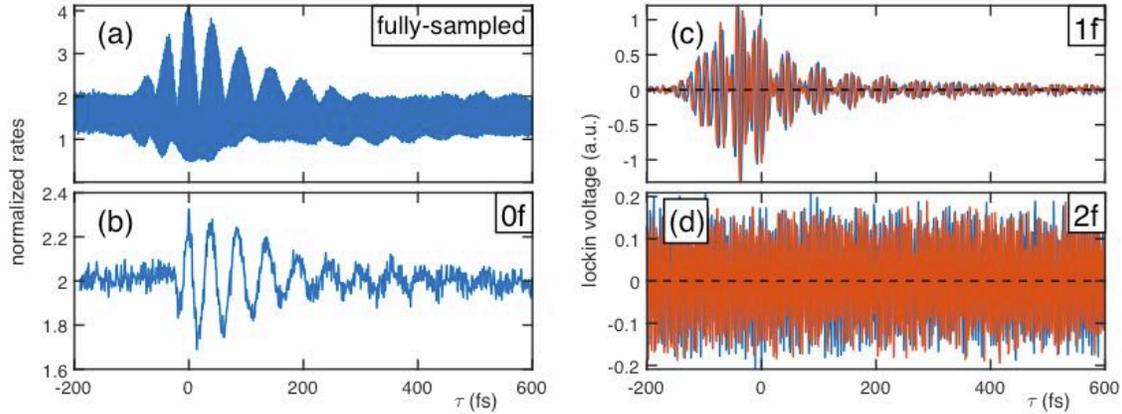

**Figure 16.** Measured two-photon coincidence interference. (a) fully sampled with 15 nm steps, (b) modulation applied without demodulation with 150 nm steps, (c) first harmonic with 150 nm steps, (d) second harmonic with 150 nm steps. Modulated data was acquired with multimode fiber-coupled APDs.

For the experiments using the notch filter, where the transmitted signal was weak, we employed multi-mode optical-fiber-coupled APDs in the collection setup. This has the benefit of essentially eliminating spectral clipping due to angular dispersion inside the MZI, at the expense of reduced spatial mode overlap at the MZI output, which resulted in reduced visibility.

Similar to the procedure with the quartz sample, we Fourier transformed the 1f component of the signal to recover the spectral features of the notch filter, as shown in **Figure 17**. The phase discontinuity is due to the lack of phase information resulting from the absence of transmitted signal in the central region. Where the signal is nonzero, the measurement follows expectation well. The recovery of the spectroscopic features of the notch filter, in qualitative agreement with expectations, demonstrates the utility of being able to address separately the 0f, 1f and 2f components of the phase-modulated two-photon coincidence rate. In the spectral regions where the notch filter transmits light, we see that the recovered phase matches qualitatively the dispersive part of the response in our model in Figure 14, as required by the Kramers-Kronig relations. The absorptive part also contains information about our sample, and due to the symmetric nature of our notch filter, qualitatively reproduces the transmission spectrum.

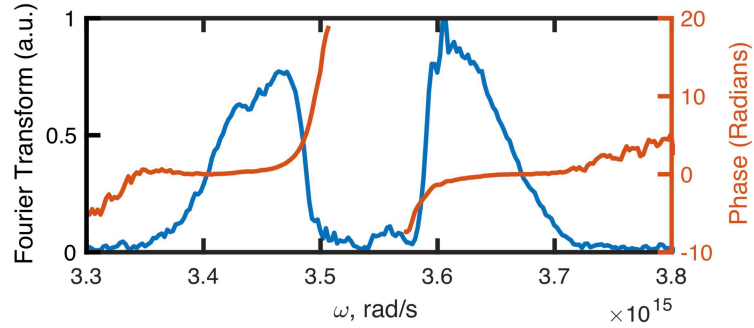

**Figure 17.** Frequency-domain magnitude and phase of the coincidence-rate interference scans with notch filter inserted. Phase corresponds to the phase of the sample. (The expected linear and quadratic phase from the notch filter substrate, fused silica, are removed.) Magnitude is a modified transmission spectrum of the sample multiplied by the EPP spectrum.

## 6. Conclusion/Discussion:

In this work, we have demonstrated a new method to use time-frequency-entangled photon pairs (EPPs) to measure the absorptive and dispersive linear response of a material sample. The technique is based on the detection of two-photon coincidences after passing through a Mach-Zehnder Interferometer (MZI) and utilizes signal isolation principles of phase-modulation and lock-in detection. The approach allows one to decompose the total coincidence signal into separate quantum optical coherence pathways of the EPP-MZI, and to understand theoretically how these component signals contribute to the total signal as a function of the relative path delay. The three component signals are associated with i) the "Hong-Ou-Mandel" (HOM) pathways, which are independent of the phase modulation and identified as the 0f signal component; ii) the "N00N" pathways, which occur at twice the phase-modulation frequency 2f; and iii) pathways created from interferences between HOM and N00N pathways, which occur at the phase-modulation frequency 1f.

We performed test experiments and theoretical calculations to characterize the properties of the EPP-MZI under three different conditions: i) in the absence of a sample, ii) with a purely dispersive sample placed in one arm of the MZI, and iii) with a complex (absorptive and dispersive) sample placed in one arm of the MZI. For all of these cases, our theoretical analyses provided simulated data that agree well with experiment and provided the necessary insight to understand how the various quantum interferences generated in the experiment are affected by the linear material response.

Of particular interest is the symmetry of the 0f HOM interference component of the signal. The term $\eta(\omega)\eta^*(2\omega_0 - \omega)$ in Equation(24) selects higher-order dispersion of only odd powers. This effect is responsible for the narrowness we observe in the HOM feature for the quartz sample shown in in Figure 12(b). This suggests that the effect can be leveraged to quantify higher-order dispersion in transparent samples where large amounts of second-order dispersion may obscure third-order dispersion.

Conversely, in the case of a monochromatic pump field, the 2f N00N-interference component of the signal contains no spectroscopic information about the sample, other than a trivial phase shift, as verified by our test experiments. When combined with the phase-modulation technique, it is possible to step the delay of the MZI with resolution below the expected Nyquist rate for the highest frequency (2f) component in the 'rotating' (down-sampled) frame, without concern for introducing aliasing artifacts that might otherwise result from interference with the N00N pathway. This further relaxes our sampling requirements by a factor of two.

Information about the sample linear response is encoded in the 1f signal component. While the absorptive part of the material response is encoded in a non-trivial manner, the dispersive part is encoded plainly, and can be recovered easily, even for challenging samples such as the notch filter for which the transmission approaches zero within the central spectral region.

The methods developed in this work are a necessary step towards the implementation of EPP spectroscopy to measure the *nonlinear* response of samples. For example, EPP-two-dimensional fluorescence spectroscopy (EPP-2DFS) has been proposed to measure the 2D nonlinear electronic spectra of coupled chromophore systems.[26]

## Acknowledgements

The authors were supported by a grant from the John Templeton Foundation, "Quantum Simulators of Complex Molecular Networks," grant number 60469, and a grant from the US NSF, "RAISE-TAQS: Quantum Advantage of Broadband Entangled Photon Pairs in Spectroscopy," grant number PHY-1839216.